# Teaching Introduction to Programming in the Times of AI: A Case Study of a Course Redesign

Nikolaos Avouris[1], Kyriakos Sgarbas[1], George Caridakis[1], Christos Sintoris[2]

avouris@upatras.gr, sgarbas@upatras.gr, gcari@aegean.gr, sintoris@upatras.gr

[1] Adjunct Professor, Hellenic Open University, Greece
[2] Laboratory Teaching Staff, University of Patras, Greece

## Abstract

The integration of AI tools into programming education has become increasingly prevalent in recent years, transforming the way programming is taught and learned. This paper provides a review of the state-of-the-art AI tools available for teaching and learning programming, particularly in the context of introductory courses. It highlights the challenges on course design, learning objectives, course delivery and formative and summative assessment, as well as the misuse of such tools by the students. We discuss ways of re-designing an existing course, re-shaping assignments and pedagogy to address the current AI technologies challenges. This example can serve as a guideline for policies for institutions and teachers involved in teaching programming, aiming to maximize the benefits of AI tools while addressing the associated challenges and concerns.

## Introduction

The rapid advancements in Artificial Intelligence (AI) have brought transformative changes across numerous fields, including education. In particular, the teaching of introductory programming—long framed around traditional lectures, guided exercises, and manual grading—is being reshaped by AI-powered tools that offer real-time feedback, automate code generation and evaluation, generate instructional content, and provide adaptive learning experiences. Tools such as GitHub Copilot, ChatGPT, and numerous AI-based tutoring systems are drastically changing how students learn to program and how instructors design, deliver, and assess their courses. This paper explores the impact of AI tools on teaching and learning of introduction to programming, outlines the corresponding shifts in educational objectives and pedagogical considerations, and proposes through a case study, a set of good practices for institutions and instructors.

The paper is structured as follows: First we review relevant literature and available tools and technologies. Then we outline the challenges in re-shaping a typical Introductory to Programming course. Next, we describe the current design and delivery of such a course and proceed with proposal for its re-design. Finally, we look into the future of institutional policies and guidelines for such transformation, aiming to support the didactics of informatics in a rapidly evolving technological landscape.

## Literature Review and Current Technological Background

Today, there is a significant number of AI tools that can support or enhance learning of introductory programming. There are tools based on AI technology for code generation and evaluation, like ChatGPT or GPT-4 (OpenAI), and many other similar tools. These have been used to answer conceptual questions, offer explanations, and help debug or refactor code through natural language interactions. Similarly, GitHub Copilot (OpenAI Codex) has been used to suggest code completions, function implementations, and even entire modules based on prompts in the editor. In addition, many experiments have been made to develop

Conversational AI and Virtual Tutors, Intelligent Tutoring Systems (ITS) that tailor exercises to learners' needs and offer hints or scaffolded guidance when students need it (Chen et al. 2022, Er et al. 2024, Finn et al. 2023, Gabbay & Cohen, 2022, Wang et al. 2023). More relevant to introductory to programming, recent work by Diamantopoulos et al. (2025) provides a concrete example of how Large Language Models, specifically GPT-4, can be leveraged to deliver formative feedback in a first-year programming course. Crucially, the authors highlight potential pitfalls, such as the risk of over- or under-grading edge cases and the need for robust grading rubrics to guide the AI's evaluation process. These findings align closely with the concerns raised in other recent studies on AI-assisted programming instruction and gradings (e.g., Golchin et al., 2025; Nagakalyani et al. 2025; Prather et al., 2023) and support the broader call for rethinking assessment strategies, academic integrity policies, and faculty training to accommodate AI-driven teaching tools.

To address these concerns, Bozkurt et al. (2024) proposed a manifesto for teaching and learning in a time of AI. In this, they advocate for critical and reflective pedagogy on AI use and role, reformulation of traditional assessment methods, disclosure of AI involvement in assignments and assessments and call for curricula that not only integrate AI as a learning tool but also prepare students to take a critical stance towards AI. These ideas are relevant to Programming courses, where there is the danger of students offloading critical thinking and problem-solving tasks and not developing fundamental required skills. As Risko and Gilbert (2016) have observed, cognitive offloading can even be beneficial to learning, if tools are used for routine tasks, enabling students to focus on higher-level problem solving and creative tasks. Such an example being support for syntactic aspects of coding. However today there is a danger of over-reliance on external aids that may lead to shallow learning, so there is a need for updated pedagogical policies that address the balance of AI benefits without undermining the deeper learning process. Along these lines, Tan et al. (2025) examine the dual role of AI in computer science education, emphasizing that current technologies are both powerful tools for enhancing programming productivity while they acknowledge the danger of students becoming overly reliant on automated solutions, without internalizing core programming concepts. They suggest adaptation of curricula that integrate AI modules and propose balanced assessment practices based on clear academic integrity policies.

Finally, Garcia (2025) conducted a review of literature on the use of ChatGPT in teaching and learning computer programming, identifying both pedagogical opportunities and challenges associated with generative AI technologies. The review concludes that while ChatGPT can support novice programmers through immediate code generation, explanation, and debugging assistance, it also risks fostering overreliance, potentially undermining the development of foundational computational thinking and problem-solving skills. Garcia highlights a shift in the instructional landscape where educators must balance AI-enhanced support with the cultivation of core programming competencies. Among the key recommendations are the integration of AI literacy into curricula, the redesign of assessments to focus on higher-order thinking, and the promotion of reflective practices to ensure students understand—not just use—code. These conclusions underscore the urgency for programming courses to be restructured to encourage computational thinking and abstraction.

## Instructional design for the times of AI

In view of the significant technological advances and the new function of technology as a productivity tool in computer science, as discussed in the previous section, we need to re-examine and adapt their instructional designs to consider the fast-changing technological background, and at the same time, prevent the misuse of these same tools, which could undermine deeper learning objectives.

We will discuss in this section a theoretical framework for addressing this challenge. Biggs (1996) introduced *constructive alignment* as an instructional design framework, emphasizing the need for Learning Outcomes, Teaching & Learning Activities (TLAs), and Assessment Strategies to be coherently aligned. Learning Outcomes articulate the knowledge, skills, or competences students should acquire; Teaching & Learning Activities are designed so that students perform tasks that help them achieve those outcomes, while Assessment Strategies measure whether and how well students have achieved these outcomes. When these three components line up, students can clearly see the connection between course activities, assessments, and the intended results, thus better engaging and learning (Figure 1).

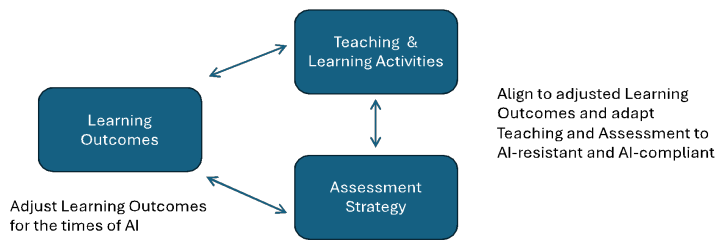

**Figure 1. Tasks of Course adaptation in the times of AI (adopted from Biggs (1996)**

The re-design interventions that need to be undertaken, as depicted in Fig.1, are relating to the components of this framework.

(a) *Adjustment of the Learning Outcomes*. We need to re-define the learning outcomes, given the existence of the new technological background and tools. We need to clearly define what the students should be able to know and skills to acquire with respect to the AI assistance. For example, we need to state clearly that students should be able to apply fundamental programming concepts independently of AI assistance, and on the other hand to define new skills relating to the existence of AI technology, like for example, that students will critically evaluate AI-generated code for correctness, efficiency, and ethics, thus emphasizing not only coding proficiency but also critical thinking and ethical considerations specific to AI.

(b) *Modify Teaching and Learning Activities*. Given the learning outcomes re-design, we need to modify teaching and learning activities. For instance, we need to include structured tasks where students use AI-assisted tools (e.g., code completion, automated debugging) under guided conditions. Activities might include comparing outputs from an AI code assistant to students' own code, thereby prompting critical reflection. We may also need to add new activities, relating to topics like ethics & policies. These align with learning outcomes addressing the responsible use of AI, by hosting group discussions or projects that examine AI's ethical implications in programming.

(c) *Adapt Assessment Strategy*. In addition, we must revise our assessment practices, so they reflect both the new learning outcomes and the activities and ensure fair use of the available technologies. For instance, we can change evaluation, so that we do not only check final code correctness but also how students arrived at their solutions. This could involve code walkthroughs, self-reflections, or AI-use disclosure statements—all aligned with the outcome of fostering critical thinking about AI outputs. We may need to adapt AI-resistant evaluation methods, like oral examinations and live coding, an example being face-to-face or recorded demonstrations to confirm authentic student understanding and mitigate overreliance on AI. In addition, if the learning outcomes include ethical/critical use of AI, we need to develop rubrics that assess whether students have appropriately employed AI tools, cited them, and provided critiques of AI-generated code.

In the following sections, we describe a case study of a proposed re-design of an Introduction to Programming course, starting with description of current course design.

## A case study: Introduction to programming course in HOU

In this section we describe the current design of an Introductory in Programming Course of HOU, Greece, and we examine scenarios for use of available AI tools by the students. The course is a distance learning course, delivered over a period of 38 weeks. The learning objectives - typical of such an introductory course - include, using Python as a programming language, to introduce students to the principles of programming for solving problems and creating real-world applications using appropriate data structures and abstraction levels. It covers various programming paradigms—procedural, object-oriented, and functional—allowing learners to select the optimal approach, verify their solutions, and document functionality effectively. Additionally, the course offers practical experience in connecting applications to databases, working with event-driven programming, and utilizing modules for tasks like data analysis, scientific computing, and machine learning, thereby highlighting the role of computing in addressing societal challenges. Currently, no specific reference to AI tools and their use is included in the learning outcomes.

The course activities include five online tutorial sessions for groups of up to 25 students, while extra one-to-one tutoring can be delivered if required. The students are asked to submit four assignments, and to work on a large project for a period of 20 weeks, in groups of 4 to 5 students. Their assessment is based on grades of the four assignments (30%) and on assessment of the project work (70%).

Each of the four assignments, features a number of questions. These assignments progressively cover all course topics—from introductory material to advanced data processing modules. Often the questions are accompanied by screenshots of expected performance, and in some cases, by code templates and additional data files.

Next, we describe the findings of a study aiming to evaluate how a contemporary AI tool would tackle recent assignments. So, sued the assignments of the academic year 2024-2025 to an AI tool and assessed its performance. The findings are included in Table 1, for the *chatGPT o3-mini-high* model (the most suitable OpenAI model for code generation).

The model answered most of the questions correctly without needing a follow-up prompt—simply by processing the question document and supplying any additional files. Notably, both the original questions in the assignment, as well as the provided answers were in Greek. The model achieved an impressive overall score of 38.4 out of 40, placing it in the top 10% of a group of currently 20 enrolled students.

However, some limitations were observed. For instance, in a simple exercise part of Q1.1: "Remove as many parentheses as possible from the following expression: *(a\*(3/b)) + ((a+2b) + (5a-b) - (a+b))*." The expected answer is *a\*\*(3/b) + a + 2b + 5a - b - (a+b)*.

Despite its apparent simplicity, this problem has proven difficult for some state-of-the-art language models. For example, both *chatGPT o3-mini-high* and *Grok-3* returned incorrect expressions, *a\*\*(3/b) + (a+2b) + (5a-b) - (a+b)*, while *Claude 3.7 Sonnet a\*\*(3/b) + 5a*, and *Gemini-2* proposed the following expression: *a\*\*(3/b) + a + 2b + 5\*a - b - a - b*.

In some other cases, the model failed to consider the associated templates or data files, e.g. an example was, not checking the character encoding of a data file. In general, however, most of the more complex questions that required generating code and integrating external files were answered correctly. The size of the code necessary to answer these questions varied between 50 and 160 lines. These responses included not only the correct code but also clear comments and a trace of the underlying problem-solving reasoning.

| question | topic | points | extras | AI score | comments |
|---|---|---|---|---|---|
| Q1.1 | Expressions, logical operators | 3 | | 2.8 | failed to eliminate a non-needed parenthesis in an expression |
| Q1.2 | Loop structures | 3 | | 3.0 | |
| Q1.3 | Data Input, Execution Flow Control | 4 | | 4.0 | |
| Q2.1 | Strings, Sets | 3 | template | 3.0 | |
| Q2.2 | Lists, Tuples | 3 | template | 3.0 | |
| Q2.3 | Dictionaries | 4 | template | 3.6 | The program had no way to exit when asked for a record ID when there were no records in the list |
| Q3.1 | Classes, Objects, Properties | 3 | | 3.0 | |
| Q3.2 | File Reading, Processing, Updating | 3 | data files | 2.5 | The data files were not encoded in UTF8 and the program did not check for encoding, failed to read. |
| Q3.3 | Classes, Overloading | 4 | template | 4.0 | |
| Q4.1 | File Management - Storing and retrieving objects with pickle | 3 | | 3.0 | |
| Q4.2 | NumPy library | 3 | | 3.0 | |
| Q4.3 | pandas, NumPy and matplotlib libraries | 4 | data file, template | 3.5 | Originally failed to use the template that was provided, after prompting, it provided full solution |
| | total | 40 | | 38.4 | |

**Table 1. Assigned work (HOU, Acad. Year 2024-25): AI tool performance**

Next, we tested the group project, which represents the most critical activity for the course, also serves as final assessment. The project requirements specify that students must deliver their code to solve a complex problem and produce a 10-page report explaining the problem, their approach, work distribution, and results. Additionally, each group member had to submit a 2-page individual report detailing their specific contribution, workload, and resources used. Students selected their group project topics from a predefined list.

To evaluate the capabilities of an AI tool in supporting student projects, we examined a case study using, as an example one of the project topics, the "Memory Game" project—a card game for 1-4 players with three difficulty levels (played with 16, 40, or 52 cards). We input the three-page project specification into the *chatGPT o3-mini-high* model, which generated over 350 lines of code. The application, following the instructions, used the *tkinter* graphics module for the graphic user interface. The code was organized into three classes (*Card*, *Player*, and *MemoryGame*).

The initial code execution revealed however, significant issues. We followed a few test cycles and iterative feedback to the AI tool; we obtained progressively improved versions until reaching a fully functional implementation. An example of this refinement process is shown through the follow-up prompt used: "*The generated code has the following problems: when two players are playing, and the cards run out, the game does not end. Also, the GUI does not strongly indicate which player's turn is to play. Also, in the case of a player against the computer it does not continue after the first move*", the result of this prompt, was a new version of the code with explanation of the modifications. This cycle of test-modify was repeated several times. Figure 2 illustrates through screenshots this evolution: version (a) featured interaction through pop-up messages; version (b) incorporated player's-turn message at the top of the window and score statistics at the bottom; and version (c) further improved the application, replacing gray button cards with colored labels.

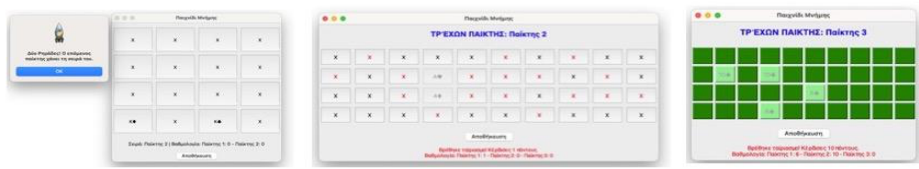

**Figure 2. Three consecutive versions of the Memory Game application**

Completing this complex task, with the help of AI, took approximately two hours, resulting in a final version comprising approximately 400 lines of code. Importantly, no Python programming was involved, and there was no discussion of the software's architecture or functionality—leaving the codebase essentially a black box.

Then, we requested assistance with the final project report that had to be produced. The AI tool generated precise report guidelines and, when prompted about the content of the individual reports of the group members, suggested the following work division: *Member 1* handling core logic and backend; *Member 2* responsible for the Graphical User Interface; *Member 3* managing testing and version control; and *Member 4* overseeing documentation and final presentation. The AI tool also provided detailed instructions for structuring the individual reports according to the specification. The obvious ethical issues relating to these suggestions are open for discussion.

In conclusion, our study revealed significant vulnerabilities in the current assignment structure when confronted with modern AI capabilities. The course—particularly in its distance learning format with minimal instructor-student interaction—appears susceptible to workarounds through AI tools. Students can potentially leverage these technologies to rapidly generate satisfactory responses even to complex tasks like the group project, undermining however the learning objectives.

It is evident from this study, that without thoughtful re-design that accounts for students' ready access to powerful AI assistants, the course risks failing to achieve its intended learning outcomes. The ease with which AI can produce solutions in minimal time presents a compelling temptation that may compromise the development of skills and knowledge the course is designed to achieve.

In the following section, we will discuss ideas for re-designing the specific course, taking into account given constraints on its form and delivery, and finally make suggestion of guidelines, for redesign it, applicable to other similar introductory to programming courses.

## Course re-design

As discussed in the previous section, the case study course, like many other similar courses, has not yet adapted its *learning outcomes* to the current AI trends. The first question to be asked is in what degree we need to re-define learning outcomes.

Despite the AI trends, the underlying objectives of introductory programming courses remain largely consistent. Typically, such courses aim to: Develop Computational Thinking (Wing, 2006), that is the ability to break down problems into programmable steps, identify patterns, and apply abstraction effectively, to develop Problem-Solving Skills which include designing, implementing, and testing algorithms using fundamental data structures and control structures. To promote understanding of core Programming Concepts: variables, data types, control flow (conditional statements, loops), functions, modules, and basic data structures (lists, arrays, dictionaries, etc.) in various programming paradigms (often extending to Object-oriented programming), understand and practice with software engineering best practices, like code readability, documentation, testing, debugging, and version control. The learning objectives may also include soft skills like encourage collaboration and communication through group work which includes peer reviews, pair programming, and effectively communicate and present programming ideas. These key learning outcomes remain relevant and are not affected by the existence of AI tools.

However, there is a growing number of researchers that have observed that this needs to be enhanced with new learning objectives, like: Understanding AI capabilities and limitations, critical engagement with AI tools, ethical aspects on use and academic integrity, data privacy and fairness, as well as prepare students for a rapidly evolving landscape (Mahon et al. 2024). So, in the case of the case study course, the proposal is to extend the learning outcomes with two new topics: (i) AI tools capabilities and limitations and (ii) Ethical aspects. Given that the existing learning outcomes already cover the expected workload, some peripheral outcomes, like databases, and data analysis modules may need to be given less attention as shown in Figure 3.

> 1. Promote understanding of core **Programming Concepts** in Python: variables, data types, control flow (conditional statements, loops), functions, modules, and basic data structures (lists, arrays, dictionaries, etc.) using various programming paradigms (Procedural, Object-oriented, Event-based),
>
> 2. Develop the ability to **break down problems** into programmable steps, identify patterns, and apply **abstraction** effectively,
>
> 3. Develop **Problem-Solving Skills** which include design, implementation, and testing of algorithms using fundamental data structures and control structures.
>
> 4. Understand and practice with **software engineering** principles,: code readability, documentation, testing, debugging, and version control.
>
> 5. ~~Get experience in connecting applications to databases, working with event-driven programming, and utilizing modules for tasks like data analysis, scientific computing, and machine learning~~
>
> 6. Develop skills of **collaboration and communication** including peer reviews, pair programming, and effectively communicate and present programming ideas.
>
> 7. **Artificial Intelligence programming tools** capabilities and limitations -  Ethical aspects

**Figure 3. Updated Learning Outcomes for the course of our case study**

The second modification concerns the *learning and teaching activities*. As discussed in the theoretical section, these need to consider the enhanced learning outcomes and on the other hand, adapt to the existence of powerful tools that the students may be tempted to use, undermining the learning process. For our case, these activities include: (a) the five online tutoring sessions, (b) the four assignments and (c) the project work.

First, we need to adjust the topics of the online tutoring sessions to accommodate the modified learning outcomes. The current practice of the tutorials foresees for each one of them a four-hour session that has the following structure: Four of them are related to the assignments, including presentation of the next assignment and discussion of the typical errors of the previous one, presentation of the related theoretical notions with typical short snippets of code provided as examples, while the fifth tutorial has slightly different structure, as it is preparation for the final exam and revision of all the topics covered during the year. The current structure allows little time for interaction with the students, who mostly passively attend the presentation. The topics of the four main tutorials are the following:
1. Introduction to Python, variables, control structures, simple data structures
2. Complex data structures, Functions - Procedural programming
3. Object-oriented programming, - File I/O operations
4. Event programming, Modules, Data analysis

As discussed, most of the topics of tutorial #4 have been proposed to be dropped, while we have introduced two new topics, (a) the AI programming tools, and (b) ethical aspects.

The new structure that accommodates these modifications is presented next.

*Tutorial #1*: Python foundation and Introduction to AI tools. We introduce basic syntax, variables, data types, control structures and we present AI code generation examples (e.g. through *GitHub Co-pilot*). We select an example that needs inspection and modification of the code generated by the AI tool and discuss the process. We introduce the discussion on the responsible use of AI tools and potential pitfalls (e.g., code bias, incorrect suggestions).

*Tutorial #2*: Complex Data Structures, Functions - Procedural programming, and Responsible AI Usage. We introduce complex data structures (list, dictionaries, etc.) and operations in them, then we introduce functions and procedural programming. Finally, we do a live demo of AI suggestions for a function implementation. We introduce the notion of academic integrity, how to disclose AI assistance and potential code plagiarism.

*Tutorial #3*: Object-Oriented Programming, File I/O operations, and AI Ethics. In this tutorial we focus on the Object-oriented programming paradigm and file I/O operations, in addition, we introduce theoretically and through examples issues related to bias and fairness in AI suggestions, present examples of verifying correctness and efficiency of AI-generated code and strategies for debugging AI outputs.

*Tutorial #4*: Modules, Advanced Python Features, and AI Integration. In this tutorial we focus on importing custom modules and widely used python libraries. We introduce simple examples of integration with AI modules (e.g. Hugging Face *transformers*). We summarize ethical guidelines, in view of the group project work, and demonstrate through a mini project integrating AI suggestions.

Finally in *Tutorial #5* we review all the topics covered and raise the discussion on the professional and societal impact of AI code.

This suggestion introduces early in the course the AI programming tools and raises gradually awareness on the pitfalls and best ethical use of them through responsible coding. It should be added that short assignments will be given during these tutorials to the students, to experiment with the tools and best practices.

Next, we need to adapt the assignments and project specification. As seen in the previous section, the current assignments and projects are not AI-resistant, while there have been no clear instructions provided on the use of available AI tools. Given the new learning outcomes and teaching activities' structure, discussed above, the assignments need to be restructured to contain both AI-resistant and AI-compatible questions while clear instructions should be given in all of them on the use of AI tools. There have been various suggestions on how to build AI-resistant assignments. A summary of them is the following (Northern Michigan University Centre for Teaching and Learning, 2025):

- During design of assignments and projects we need to test beforehand with AI tools and modify them accordingly.
- Use personalization, tailor assignments and questions to individual students and current events
- Include reflective components. Ask the students to include reflection on their answer, to document challenges they faced and how they overcame them.
- Process-based assessment, this is particularly relevant to the project, where we should ask the students to present and evaluate the progress of their work in stages.

An example of modification of one of our current questions is the following:

*Original Question* (Q1.2 2024-2025): *Develop a Python program that asks from the user a number between 1 and 10 and prints the multiplication table for this number. Clear specification was given on* *the typical input and output*. NB. This question was answered with no flows by the AI tool.

*Modified Question*: *Develop a Python program that salutes you with your name and asks you for a number between 1 and 10 and prints the multiplication table for this number, decorated accordingly. Use your imagination for presenting a nicely decorated table*. NB. Once this version was given to an AI tool, the answer included a loop, asking the user his/her name, that is not what was requested, while the decoration part was tackled through two mundane borders of 25 stars at the top and bottom of the table, so this is not anymore, a full marks solution.

In addition, on all assignments, in line with the course theoretical stance towards AI tools use should include clear instructions. An example is the following.

*In this programming course, we encourage you to explore and utilize AI tools to enhance your learning and problem-solving process. However, maintaining academic integrity requires complete transparency about how these tools are used in your submitted work. Therefore, you must explicitly acknowledge the use of any AI tools that contributed to your code, explanations, or problem-solving approach in the following manner: At the beginning of each submitted programming assignment file (e.g., .py file), you must include a dedicated section titled "AI Assistance Declaration". This section should appear as a block comment at the very top of your code file. Within this section, you must provide the following information for each AI tool you utilized in the development of that specific assignment: Name of the AI Tool: Specify the exact name of the AI tool used (e.g., ChatGPT, Gemini, GitHub Copilot). Tool Version information, Specific Ways the AI Tool Was Used: Clearly and concisely describe the specific tasks for which you used the AI tool. Be precise and avoid vague statements. Include the specific prompt used. "Generated initial code structure for the calculate_average function using prompt...."*

The final task we need to tackle is related to the assessment strategy. Since assessment in this course is based on the four assignments and the presentation of project work, we modify both further in the way they are assessed. For the assignments we introduce an oral presentation of selected parts of the submitted work. This takes the form of short questions on specific parts of the answers, especially if there are doubts on academic integrity of the provided answers. Finally, the project should follow a process-based assessment, that the groups must present and report on progress in stages, with special focus on challenges they faced and how they overcame them. This not only encourages deeper engagement but also makes it more challenging to use AI tools dishonestly.

## Conclusions and further work

This paper investigates the pedagogical challenges confronting course designers of introductory programming curricula in light of advancements in artificial intelligence. A specific course has been examined as a case study, revealing limitations inherent in current instructional methodologies. Based on this analysis, we proposed a series of modifications intended to mitigate the potential for AI tools to negatively impact the acquisition of core programming and computational thinking competencies, particularly if students adopt a passive stance towards AI-generated code. The need to acknowledge the mixed impact of these tools, capable of enhancing efficiency yet simultaneously posing a risk to fundamental skill development, is underscored. By updating learning ourcomes, embracing innovative assessment methods, and articulating clear policies on permissible AI usage, instructors can preserve the integrity of learning while harnessing the benefits of AI.

The study faces several limitations, most notably the absence of evidence regarding the effectiveness of the proposed redesign, which has yet to be approved and implemented. The next steps involve institutional-level discussion and adoption of the proposal, along with investment in faculty development and appropriate infrastructure. Through these efforts, we aim to better equip our institutions to educate a new generation of computer scientists—individuals who not only excel in the technical aspects of coding but also grasp its ethical, creative, and conceptual dimensions.

## References

Biggs, J. (1996). Enhancing teaching through constructive alignment. *Higher education, 32*(3), 347-364.

Bozkurt, A., Xiao, J., Farrow, R., Bai, J. Y., Nerantzi, C., Moore, S., ... & Asino, T. I. (Eds.). (2024). The manifesto for teaching and learning in a time of generative AI: A critical collective stance to better navigate the future. *Open Praxis, 16*(4), 487-513.

Chen, X., Xie, H., Zou, D., & Hwang, G.-J. (2022). Application and impact of AI in education. *Educational Technology & Society, 25*(1), 1-15.

Diamantopoulos, A., Sintoris, C., Demetriadis, S., Avouris, N. (2025). Feedback to Students and Instructors of a Programming Course through a Large Language Model, in *Proceedings 14th Panhellenic Conference iCT in Education*, Rhodes, October 2025.

Er, E., Akçapınar, G., Bayazıt, A., Noroozi, O., & Banihashem, S. K. (2024). Assessing student perceptions and use of instructor versus AI-generated feedback. *British Journal of Educational Technology*, 56(3), 1074-1091.

Finn, A., Petre, M., & Remy, S. (2023). Balancing AI assistance and student agency in introductory programming courses. *ACM Transactions on Computing Education, 23*(2), 14-28.

Gabbay, H., & Cohen, A. (2022). Exploring the Connections Between the Use of an Automated Feedback System and Learning Behavior in a MOOC for Programming. In *EC-TEL 2022 (LNCS 13450*, pp. 116–130). Springer. https://doi.org/10.1007/978-3-031-16290-9_9

Garcia, M. B. (2025). Teaching and learning computer programming using ChatGPT: A rapid review of literature amid the rise of generative AI technologies. *Education and Information Technologies*, *2025*, 1-25.

Golchin, S., Garuda, N., Impey, C., & Wenger, M. (2025). Grading massive open online courses using large language models. *Proceedings of the 31st International Conference on Computational Linguistics (COLING)*. Association for Computational Linguistics, arXiv preprint arXiv:2406.11102..

Mahon, J., Mac Namee, B., & Becker, B. A. (2024). Guidelines for the evolving role of Generative AI in introductory programming based on emerging practice. *Proceedings of the 2024 on Innovation and Technology in Computer Science Education*, v. 1, pp. 10-16, ACM Publication.

Nagakalyani, G., Chaudhary, S., Apte, V., Ramakrishnan, G., & Tamilselvam, S. (2025). Design and evaluation of an ai-assisted grading tool for introductory programming assignments: An experience report. *Proceedings of the 56th ACM Technical Symposium on Computer Science Education* (v. 1, pp. 805-811). ACM.

Northern Michigan University Centre for Teaching and Learning (2025). Creating AI-Resistant Assignments, Activities, and Assessments (Designing Out). Available from https://nmu.edu/ctl/

Prather, J., Denny, P., Leinonen, J., Becker, B. A., Albluwi, I., Craig, M., Keuning, H., Kiesler, N., Kohn, T., Luxton-Reilly, A., MacNeil, S., Petersen, A., Pettit, R., Reeves, B. N., & Savelka, J. (2023). The robots are here: Navigating the generative AI revolution in computing education. In *Proceedings of the 2023 Conference on Innovation and Technology in Computer Science Education (ITiCSE 2023)*, Vol. 2, July 8–12, 2023, Turku, Finland (pp. 1–51). ACM. https://doi.org/10.1145/3587103.3594206

Risko, E. F., & Gilbert, S. J. (2016). Cognitive offloading. *Trends in Cognitive Sciences, 20*(9), 676-688.

Tan, C. W., Khan, M. A. M., & Yu, P. D. (2024). AI-assisted programming and AI literacy in computer science education. *Effective practices in AI literacy Education: Case studies and reflections* (pp. 189-198). Emerald Publishing.

Wang, Q., & Tsai, C.-C. (2023). Integration of Large Language Models in Computer Science Education: Challenges and opportunities. *British Journal of Educational Technology, 54*(1), 215-233.

Wing, J. M. (2006). Computational thinking. *Communications of the ACM, 49*(3), 33-35.